\documentclass[a4paper,11pt]{article}
\usepackage{jinstpub}
\usepackage{lineno}
\usepackage{multicol}
\usepackage[separate-uncertainty=true,range-phrase=--,range-units=single,per-mode=symbol]{siunitx}
\usepackage[version=4]{mhchem}
\newcommand\blfootnote[1]{
    \begingroup
    \renewcommand\thefootnote{}\footnote{#1}
    \addtocounter{footnote}{-1}
    \endgroup
}

\title{\boldmath The DEAP-3600 liquid argon optical model and NEST updates}

\author{S. Westerdale (on behalf of the DEAP-3600 and NEST collaborations)}
\affiliation{Department of Physics and Astronomy, University of California, Riverside, CA 92521, USA}
\emailAdd{shawn.westerdale@ucr.edu}

\abstract{As liquid argon (LAr) detectors are made at progressively larger sizes, accurate models of LAr optical properties become increasingly important for simulating light transport, understanding signals, and developing analyses. 
The refractive index, group velocity, and Rayleigh scattering length are particularly important for vacuum ultraviolet (VUV) and visible photons in detectors with diameters much greater than one meter. 
While optical measurements in the VUV are sparse, recent measurements of the group velocity of 128 nm photons in LAr provide valuable constraints on these parameters. 
These calculations are further complicated by the dependence of optical parameters on thermodynamic properties that might fluctuate or vary throughout the argon volume. 
This manuscript presents the model used by DEAP-3600, a dark matter direct detection experiment at SNOLAB using a 3.3 tonne LAr scintillation counter. 
Existing data and thermodynamic models are synthesized to estimate the wavelength-dependent refractive index, group velocity, and Rayleigh scattering length within the detector, and parameters' uncertainties are estimated. 
This model, along with {\it in situ} measurements of LAr scintillation properties, is benchmarked against data collected in DEAP-3600, providing a method for modeling optical properties in large LAr detectors and for propagating their uncertainties through downstream simulations.
Updates are also presented of the Noble Element Simulation Technique (NEST) software, widely used to model scintillation and ionization signals in argon- and xenon-based detectors.}

\keywords{Noble liquid detectors, Detector modeling and simulations, Large detector systems for particle and astroparticle physics, Simulation methods and programs }

\arxivnumber{2312.07712}

\begin{document}
\maketitle
\flushbottom
\section{Introduction}

Optical simulations are used to understand detector data, aiding the development of background and signal models and selection cuts.
As detectors grow in size and analyses increase in complexity, modeling photon propagation is increasingly important.
Liquid argon (LAr) detectors become highly sensitive above \SI{1}{\meter} sizes, as Rayleigh scattering affects photon transportation.
These effects are important for current and future detectors like DEAP-3600~\cite{amaudruzDesignConstructionDEAP36002019a}, DarkSide-20k~\cite{aalsethDarkSide20k20Tonne2018}, and DUNE~\cite{dunecollaborationDUNEFarDetector2018}.

\subsection{The DEAP-3600 detector}
\begin{figure}[htb]
    \centering
    \includegraphics[width=0.3\linewidth]{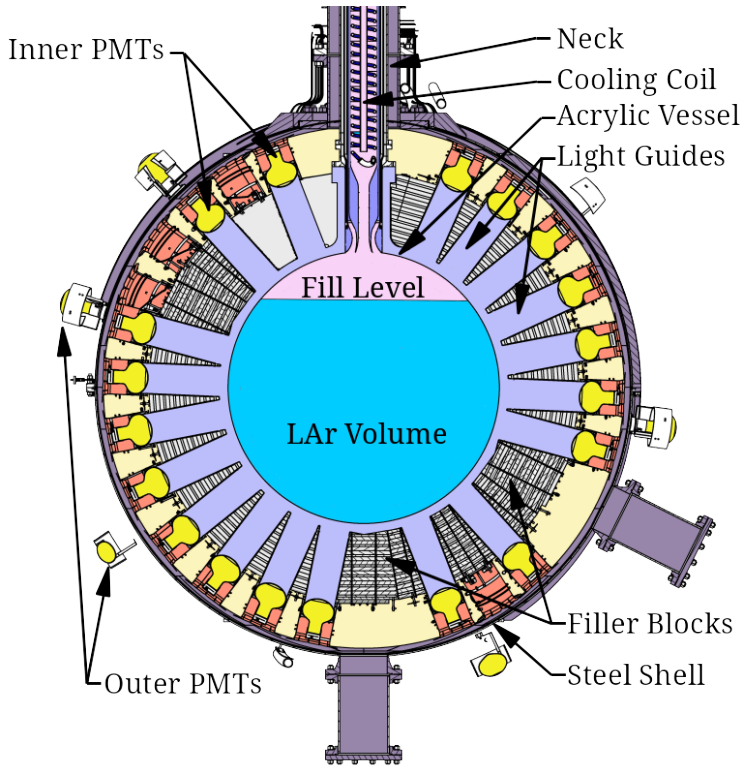}
    \includegraphics[width=0.6\linewidth]{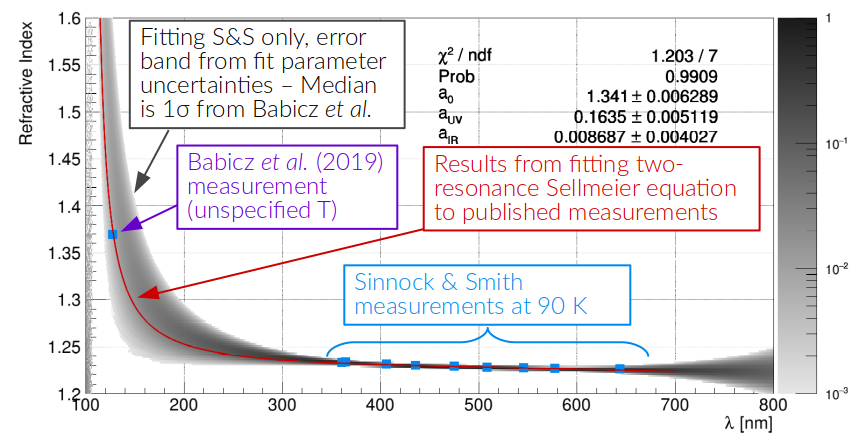}
    \caption{(Left) A cross-sectional drawing of DEAP-3600, located inside a water Cherenkov muon veto (not shown). (Right) Refractive index measurements at \SI{90}{\kelvin} fit by the Sellmeier equation (see Eq.~\ref{eq:sellmeier}).}
    \label{fig:rindexVsWavelength}
    \label{fig:deap_dwg}
\end{figure}
DEAP-3600 is a dark matter (DM) direct detection experiment operating at SNOLAB, in Sudbury, Canada, since August~2016~\cite{amaudruzDesignConstructionDEAP36002019a}.
As shown in Fig.~\ref{fig:deap_dwg}, it views \SI{3269\pm24}{kg} LAr~\cite{adhikariPrecisionMeasurementSpecific2023a} in a \SI{170}{\cm}-inner diameter acrylic vessel (AV) with an array of \num{255} photomultiplier tubes (PMTs), via \SI{45}{\cm}-long acrylic light guides, in a stainless steel shell within a water Cherenkov muon veto.
To detect VUV LAr scintillation light, the AV's inner surfaces are coated with tetraphenyl butadiene (TPB), a wavelength shifter that re-emits photons around \SI{420}{\nm}.
For more detail, see Ref.~\cite{amaudruzDesignConstructionDEAP36002019a}.

Atmospheric argon has \SI{0.964\pm0.024}{\becquerel\per\kg} of \ce{^39Ar}~\cite{adhikariPrecisionMeasurementSpecific2023a,deapcollaborationElectromagneticBackgroundsPotassium422019}.
Its \SI{565}{\keV}-endpoint $\beta$-decay provides a large dataset, used to model light production and propagation~\cite{adhikariLiquidargonScintillationPulseshape2020,adhikariPulseshapeDiscriminationLowenergy2021}.
Pulse shape discrimination---which separates electronic and nuclear recoils---and its detailed optical model allowed DEAP-3600 to constrain weakly interacting massive particles in a \SI{231}{day} exposure~\cite{deapcollaborationSearchDarkMatter2019}.
It also explored the impact of halo substructures on constraints for DM couplings in a non-relativistic effective field theory~\cite{deapcollaborationConstraintsDarkMatternucleon2020} and searched for ultra-heavy, multiple-scattering DM~\cite{deapcollaborationFirstDirectDetection2022}.
DEAP-3600's sensitivity was enabled by its optical model, used for signal and background models and cuts.

\section{LAr Optical Model}
Simulating signals in large LAr detectors requires photon production and propagation models.
For propagation, the most important properties are the refractive index, group velocity, and Rayleigh scattering length, which are all temperature-dependent and must be modeled over a wide range of wavelengths, spanning the \SI{127\pm3}{\nm} VUV scintillation wavelength~\cite{heindlScintillationLiquidArgon2010} to the \SIrange{350}{500}{\nm} wavelength range of TPB fluorescence.
Photon absorption in LAr is largely a result of impurities, and therefore varies between individual detectors. A general model is therefore not presented here.

\subsection{Modeling photon propagation}
The Rayleigh scattering length and group velocity in LAr can be derived from its wavelength-dependent refractive index and thermodynamic properties.
As in Refs.~\cite{graceIndexRefractionRayleigh2015,babiczMeasurementGroupVelocity2020}, the refractive index $n(\lambda)$ may be modeled as a function of wavelength $\lambda$ using the Sellmeier equation, approximated as,
\begin{equation}
    n^2(\lambda) = a_0 + \frac{a_\text{UV}\lambda^2}{\lambda^2-\lambda^2_\text{UV}} + \frac{a_\text{IR}\lambda^2}{\lambda^2-\lambda^2_\text{IR}},
    \label{eq:sellmeier}
\end{equation}
considering the nearest ultraviolet (UV) and infrared (IR) resonances in atomic argon, and effectively accounting for farther resonances in $a_0$.
In Eq.~\ref{eq:sellmeier}, $\lambda_\text{UV (IR)}$ and $a_\text{UV (IR)}$ are the wavelength and intensity of the UV (IR) resonance, measured using emission and absorption lines in atomic argon to be $\lambda_\text{UV}=\SI{106.7}{\nm}$~\cite{laneArgonResonanceLine2003} and $\lambda_\text{IR}=\SI{960}{\nm}$~\cite{araiInfraredAbsorptionsNeon1978}.
As in Ref.~\cite{graceIndexRefractionRayleigh2015}, $a_0$, $a_\text{UV}$, and $a_\text{IR}$ are determined as a function of temperature by fitting Eq.~\ref{eq:sellmeier} to measured refractive indices at different temperatures.

The group velocity $v_g$ at frequency $\omega$, wave number $k=2\pi/\lambda$, and phase velocity $v_p=c/n$ is,
\begin{gather}
    \begin{aligned}
        v_g(\lambda) = \frac{\partial\omega}{\partial k} = \frac{c}{n} - \frac{c}{n^2}\frac{\partial n}{\partial k}k = \frac{c}{n}\left(1+\frac{\lambda}{n}\frac{\partial n}{\partial\lambda}\right),
    \end{aligned}
\end{gather}
with vacuum speed of light $c$.
The Rayleigh scattering length at temperature $T$ is (see Refs~\cite{landau1984electrodynamics,seidelRayleighScatteringRaregas2002}),
\begin{gather}
    \begin{aligned}
        L^{-1}_\text{Rayleigh} &= \frac{\omega^4}{6\pi c^4}\left[k_BT\rho^2(T)\kappa_T(T)\left(\frac{\partial\epsilon}{\partial\rho}\right)_T^2 + \frac{k_BT^2}{\rho(T) c_V(T)}\left(\frac{\partial\epsilon}{\partial T}\right)_\rho^2\right], \\
        &\approx \frac{8\pi^3}{3n(T)^4\lambda^4}\left[k_BT\rho^2(T)\kappa_T(T)\left(\frac{\partial n^2}{\partial\rho}\right)_T^2\right],
    \end{aligned}
\end{gather}
where $k_B$ is Boltzmann's constant, $\rho$ density, $\kappa_T$ isothermal compressability, $\epsilon=n^2$ dielectric constant, and $c_V$ heat capacity at constant volume. 
Ref.~\cite{seidelRayleighScatteringRaregas2002} shows that the  $\left(\frac{\partial\epsilon}{\partial T}\right)^2$ term is typically much smaller than the $\left(\frac{\partial\epsilon}{\partial\rho}\right)^2$ term in noble liquids and can be neglected. 
The Lorentz-Lorenz~\cite{lorentzUeberAnwendungSatzes1881} and Clausius-Mossotti~\cite{clausiusMechanischeBehandlungElectricitaet1879} equations relate the refractive index to the density, giving,
\begin{gather}
    \begin{aligned}
        \frac{n^2-1}{n^2+2} &= \frac{4\pi}{3}\frac{N_A\alpha_0}{M\epsilon_0}\rho(T) = A\rho(T), \\
        \frac{\partial n^2}{\partial\rho} &= \frac{A}{3}\left(n^2+2\right)^2 = \frac{1}{3\rho}\left(n^2-1\right)\left(n^2+2\right),
        \\
        \frac{\partial n}{\partial T} &= \frac{3A}{2n}\frac{1}{\left(1-A\rho\right)^2}\frac{\partial\rho}{\partial T} = \frac{3}{2n\rho}\left(\frac{n^2-1}{n^2+2}\right)\left(1-\frac{n^2-1}{n^2+2}\right)^{-2}\frac{\partial\rho}{\partial T},
        \label{eq:clausiusmossotti}
    \end{aligned}
\end{gather}
with Avogadro number $N_A$,  electric constant $\epsilon_0$,  polarizability $\alpha_0$, constant $A$, and molar mass $M$.

The isothermal compressability is $\kappa_T(T)=\frac{1}{\rho}\left(\frac{\partial\rho}{\partial p}\right)_T$, at pressure $p$.
It is calculated from the equation of state, fit for argon (as well as neon and xenon) to within \SI{0.1}{\percent} of all data~\cite{rabinovichThermophysicalPropertiesNeon1987}, giving,
\begin{gather}
\begin{aligned}
    p(T) &= A_p(T)\rho^2 + B_p(T)\rho^4 + C_p(T)\rho^6, \\
    A_p(T) &= -241.8 + 3.8T - 330.3\times10^4T^{-2}+54.5\times10^3T^4, \\
    B_p(T) &= -192.3+2.8T+0.01T^2, \qquad
    C_p(T) = 174.7+0.9T, \\
    \kappa_T(T) &= \frac{1}{\rho}\left(\frac{\partial\rho}{\partial p}\right)_T = \left(2A_p(T)\rho^2+4B_p(T)\rho^4+6C_p(T)\rho^6\right)^{-1},
\end{aligned}
\end{gather}

Lastly, the density of LAr can be calculated using the empirical equation in Ref.~\cite{ctx43101231760006531},
\begin{gather}
    \begin{aligned}
        \rho(T) &= \rho_c\exp{\left(A_\rho\left(1-\frac{T}{T_c}\right)^{1/3}+B_\rho\left(1-\frac{T}{T_c}\right)^{2/3} + C_\rho\left(1-\frac{T}{T_c}\right)^{7/3} + D_\rho\left(1-\frac{T}{T_c}\right)^4 \right)},
    \end{aligned}
\end{gather}
with $\rho_c=\SI{0.5356}{\gram\per\cubic\cm}$, $T_c=\SI{150.687}{\kelvin}$, $A_\rho=\num{1.5}$, $B_\rho=\num{-0.31}$, $C_\rho=\num{0.086}$, and $D_\rho=\num{-0.041}$.

\subsection{Fitting measured refractive indices}
\begin{table}[htb]
    \centering
    \caption{Summary of LAr refractive indices reported in Ref.~\cite{sinnockRefractiveIndicesCondensed1969}, as a function of wavelength and temperature}
    \begin{tabular}{c|ccccccccc}\hline\hline
        & \multicolumn{9}{c}{Wavelength [nm]} \\
        T [\si{\degree\kelvin}] & 643.9 & 578.0 & 546.1 & 508.6 & 475.3 & 435.8 & 406.3 & 365.0 &  361.2  \\\hline
        83.8 & 1.2321 & 1.2328 & 1.2334 & 1.2341 & 1.2349 & 1.2361 & 1.2372 & 1.2392 & 1.2395 \\
        86 & 1.2295 & 1.2303 & 1.2308 & 1.2316 & 1.2324 & 1.2336 & 1.2347 & 1.2367 & 1.2370 \\
        88 & 1.2274 & 1.2282 & 1.2287 & 1.2295 & 1.2303 & 1.2315 & 1.2326 & 1.2346 & 1.2349 \\ 
        90 & 1.2256 & 1.2264 & 1.2269 & 1.2277 & 1.2285 & 1.2297 & 1.2308 & 1.2331 & 1.2326 \\\hline\hline
    \end{tabular}
    \label{tab:sinnock}
\end{table}
The temperature dependence of the refractive index in liquid argon, krypton, and xenon is explored for visible wavelengths in Ref.~\cite{sinnockRefractiveIndicesCondensed1969}.  LAr refractive indices are summarized in Table~\ref{tab:sinnock}.
\begin{figure}
    \centering
    \includegraphics[width=0.495\linewidth]{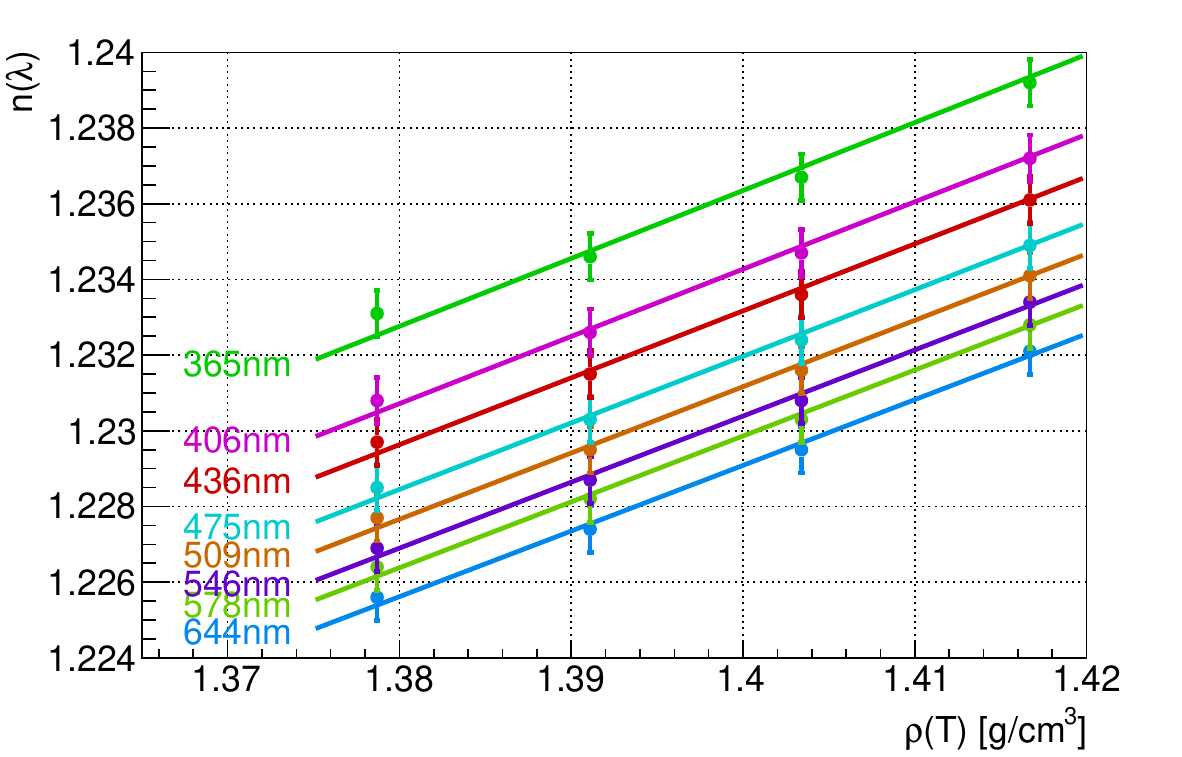}
    \includegraphics[width=0.495\linewidth]{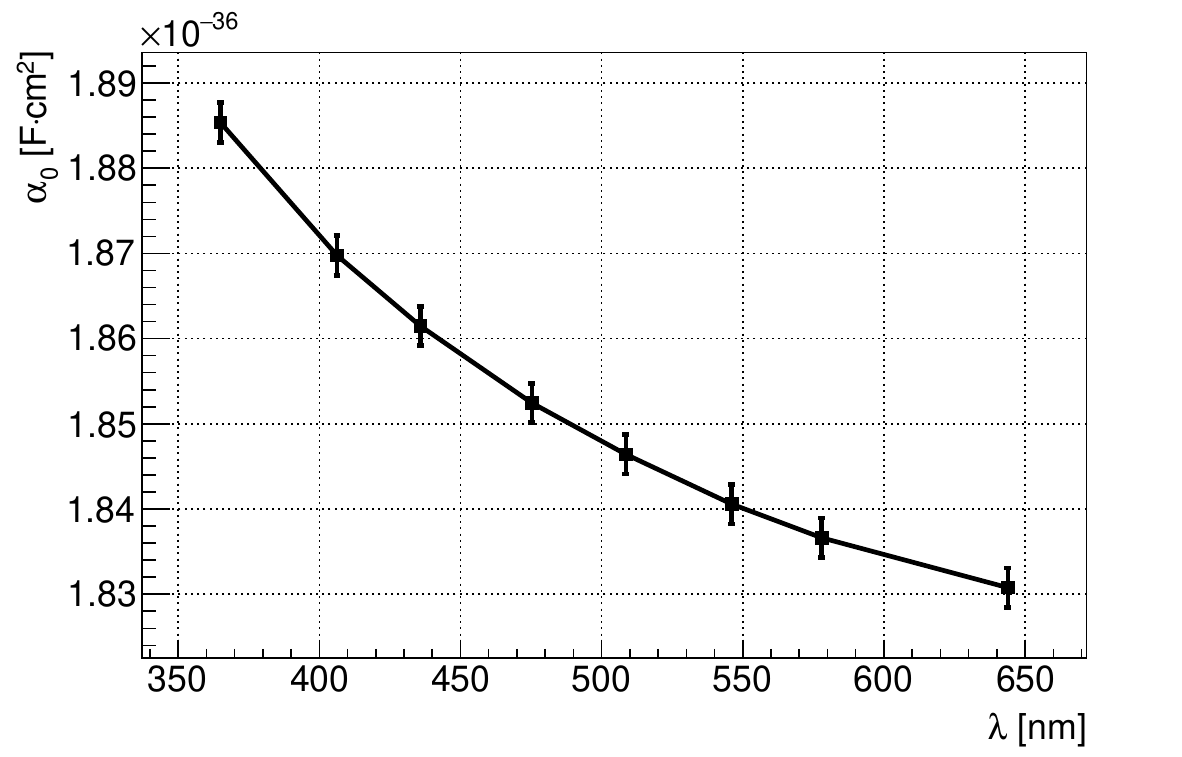}
    \caption{(Left) Visible refractive index~\cite{sinnockRefractiveIndicesCondensed1969} vs. LAr density at varying temperatures and wavelengths, fit by Eq.~\ref{eq:clausiusmossotti} with $A$ a fit parameter. (Right) Atomic polarizability $\alpha_0$, from $A$ (see Eq.~\ref{eq:clausiusmossotti}), vs. wavelength}
    \label{fig:rindexDeps}
\end{figure}
Figure~\ref{fig:rindexDeps} shows the temperature-dependent refractive indices and density, fit by Eq.~\ref{eq:clausiusmossotti}, with $A$ treated as a fit parameter. The goodness of these fits confirms that $A$ is independent of temperature. The right-hand plot in Fig.~\ref{fig:rindexDeps} shows how $\alpha_0$ derived from $A$ varies with wavelength.

Ref.~\cite{babiczMeasurementGroupVelocity2020} measures the group velocity to be $v_g = \SI{13.40\pm0.15}{\cm\per\ns}$ for \SI{127\pm3}{\nm} LAr scintillation photons at \SIrange{88}{90}{\kelvin}, using a time-of-flight measurement with tagged muons.
Figure~\ref{fig:rindexVsWavelength} shows the nominal refractive index derived from this measurement plotted alongside the visible indices from Ref.~\cite{sinnockRefractiveIndicesCondensed1969} at \SI{90}{\kelvin}. 
The gray band illustrates Eq.~\ref{eq:sellmeier} fit to the visible wavelengths, only, with best-fit parameters varied according to the fit's covariance matrix.
This curve shows that the new measurement at \SI{127}{\nm} is approximately \SI{1}{\sigma} away from the best-fit line to the visible-only fit.
Varying the refractive index at \SI{127}{\nm} and re-calculating $v_g$ at \SIrange{88}{90}{\kelvin} defines the range of refractive indices consistent with the measured value of $v_g$ at \SI{127}{\nm} to be $n=\num{1.364\pm0.005}$.
\begin{figure}
    \centering
    \includegraphics[width=\linewidth]{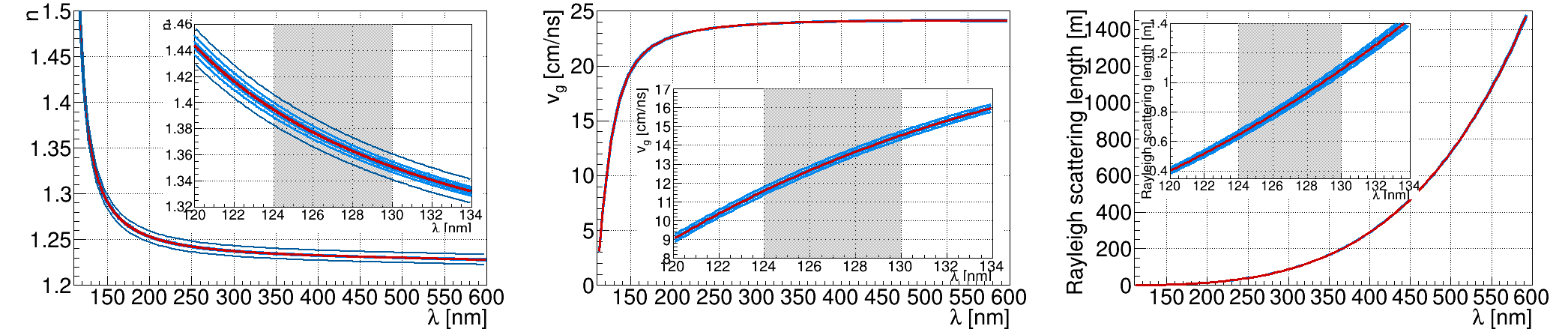}
    \caption{(Left) Group velocity, (Middle) Rayleigh scattering length, and (Right) refractive index as functions of wavelength. (Top) Zoomed in near VUV light and (bottom) showing the VUV-visible range.}
    \label{fig:vgAndRayleigh}
\end{figure}

The best-fit refractive index and corresponding group velocity and Rayleigh scattering length are shown in Fig.~\ref{fig:vgAndRayleigh}, with a band showing the effects of fit uncertainties and \SIrange{84}{90}{\kelvin} temperature variations. 
Thermal uncertainties $\delta T$ are calculated as $\sigma_T = \left(\frac{\partial n}{\partial T}\right)\delta T$.

\subsection{Modeling scintillation}
High-statistics fits to \ce{^39Ar} $\beta$-decays allow the scintillation pulse shape to be precisely measured, accounting for additional delays from TPB fluorescence and PMT afterpulses, as described in Ref.~\cite{adhikariLiquidargonScintillationPulseshape2020}.
These studies show that the pulse-shape is best described by a two-component exponential model (a fast singlet component with a \SI{<10}{\ns} lifetime and a slower triplet component with a \SI{1435}{\ns} lifetime), with an additional power law component spanning intermediate times.

\section{Validation in DEAP-3600}
\begin{figure}
    \centering
    \includegraphics[width=0.54\linewidth]{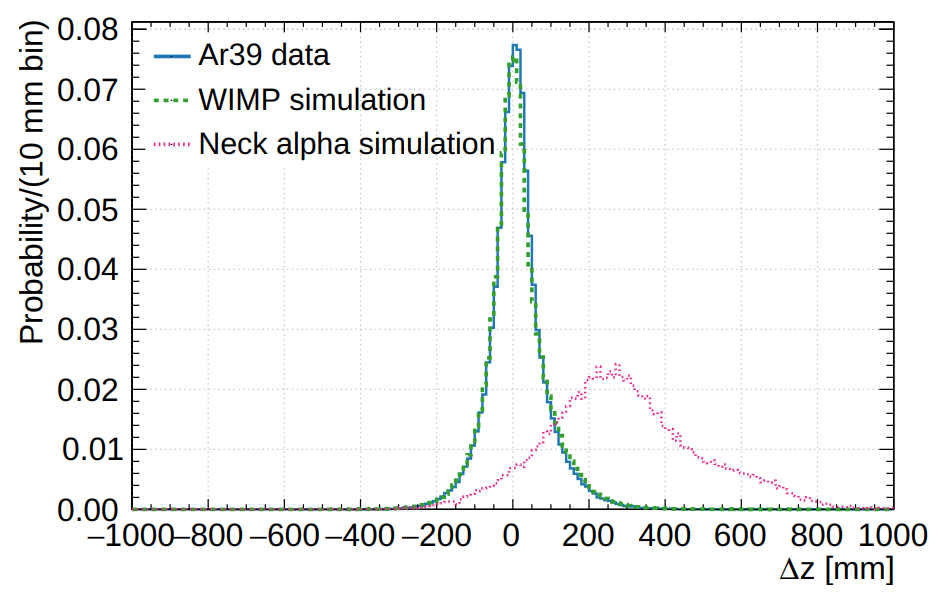}
    \includegraphics[width=0.37\linewidth]{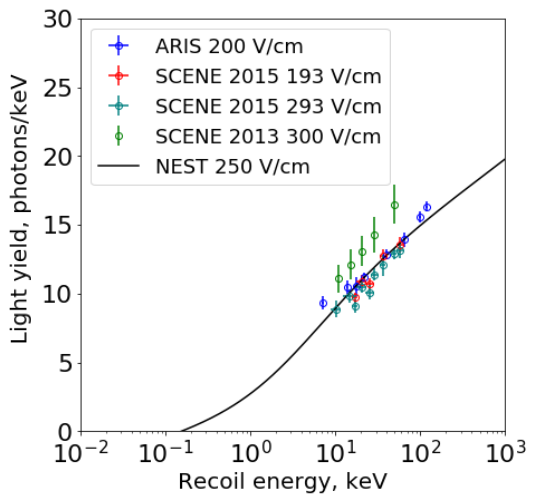}
    \caption{(Left) Comparison of the difference in the vertical position reconstructed by two algorithms, showing strong agreement between observed \ce{^39Ar} decays and simulations of WIMP nuclear recoils, expected to have the same distribution; $\alpha$-decays in the detector neck differ as expected~\cite{deapcollaborationSearchDarkMatter2019}. (Right) NEST nuclear recoil photon yield predictions~\cite{szydagisReviewBasicEnergy2021} compared to  SCENE~\cite{scenecollaborationObservationDependenceDrift2013,scenecollaborationMeasurementScintillationIonization2015} and ARIS~\cite{agnesMeasurementLiquidArgon2018} measurements near \SI{250}{\volt\per\cm}.}
    \label{fig:validation}
\end{figure}
The left panel of Fig.~\ref{fig:validation} compares data and a Geant4~\cite{rat,GEANT4} optical simulation implementing this model.
The strong agreement illustrated between high-level variables like the difference between two position reconstruction algorithms, which are highly sensitive to the optical model, validates the presented model.
Further validation, showing strong agreement between observed and simulated average waveforms and other high-level variables, is presented in Refs.~\cite{deapcollaborationSearchDarkMatter2019,sinaThesis}.

\section{NEST updates}
Separately, the Noble Elements Simulation Technique (NEST) has been developed to simulate scintillation and ionization in xenon and argon detectors.
Recent developments accurately predicted the energy resolution in XENON1T~\cite{szydagisInvestigatingXENON1TLowEnergy2021,szydagisReviewBasicEnergy2021}.
Additionally, significant progress has been made modeling the scintillation and ionization yield in LAr detectors, including strong fits to calibration measurements reported in Refs.~\cite{scenecollaborationMeasurementScintillationIonization2015,agnesMeasurementLiquidArgon2018,joshiFirstMeasurementIonization2014}, some of which are shown in the right panel of Fig.~\ref{fig:validation}.

\blfootnote{{\bf Acknowledgements:} We thank the Natural Sciences and Engineering Research Council of Canada (NSERC),
the Canada Foundation for Innovation (CFI),
the Ontario Ministry of Research and Innovation (MRI), 
and Alberta Advanced Education and Technology (ASRIP),
the University of Alberta,
University of California Riverside,
Carleton University, 
Queen's University,
the Canada First Research Excellence Fund through the Arthur B.~McDonald Canadian Astroparticle Physics Research Institute,
Consejo Nacional de Ciencia y Tecnolog\'ia Project No. CONACYT CB-2017-2018/A1-S-8960, 
DGAPA UNAM Grants No. PAPIIT IN108020 and IN105923, 
and Fundaci\'on Marcos Moshinsky,
the European Research Council Project (ERC StG 279980),
the UK Science and Technology Facilities Council (STFC) (ST/K002570/1 and ST/R002908/1),
the Leverhulme Trust (ECF-20130496),
the Russian Science Foundation (Grant No. 21-72-10065),
the Spanish Ministry of Science and Innovation (PID2019-109374GB-I00) and the Community of Madrid (2018-T2/ TIC-10494), 
the International Research Agenda Programme AstroCeNT (MAB/2018/7)
funded by the Foundation for Polish Science (FNP) from the European Regional Development Fund,
and the European Union's Horizon 2020 research and innovation program under grant agreement No 952480 (DarkWave).
Studentship support from
the Rutherford Appleton Laboratory Particle Physics Division,
STFC and SEPNet PhD is acknowledged.
We thank SNOLAB and its staff for support through underground space, logistical, and technical services.
SNOLAB operations are supported by the CFI
and Province of Ontario MRI,
with underground access provided by Vale at the Creighton mine site.
We thank Vale for their continuing support, including the work of shipping the acrylic vessel underground.
We gratefully acknowledge the support of the Digital Research Alliance of Canada,
Calcul Qu\'ebec,
the Centre for Advanced Computing at Queen's University,
and the Computational Centre for Particle and Astrophysics (C2PAP) at the Leibniz Supercomputer Centre (LRZ)
for providing the computing resources required to undertake this work.
}

\bibliographystyle{deap}
\bibliography{biblio}

\begin{thebibliography}{31}
\expandafter\ifx\csname natexlab\endcsname\relax\def\natexlab#1{#1}\fi
\expandafter\ifx\csname bibnamefont\endcsname\relax
  \def\bibnamefont#1{#1}\fi
\expandafter\ifx\csname bibfnamefont\endcsname\relax
  \def\bibfnamefont#1{#1}\fi
\expandafter\ifx\csname citenamefont\endcsname\relax
  \def\citenamefont#1{#1}\fi
\expandafter\ifx\csname url\endcsname\relax
  \def\url#1{\texttt{#1}}\fi
\expandafter\ifx\csname urlprefix\endcsname\relax\def\urlprefix{URL }\fi
\providecommand{\bibinfo}[2]{#2}
\providecommand{\eprint}[2][]{\url{#2}}

\bibitem[{\citenamefont{Amaudruz
  et~al.}(2019)}]{amaudruzDesignConstructionDEAP36002019a}
\bibinfo{author}{\bibfnamefont{P.~A.} \bibnamefont{Amaudruz}}
  \bibnamefont{et~al.} (\bibinfo{collaboration}{DEAP Collaboration}),
  \href{https://doi.org/10.1016/j.astropartphys.2018.09.006}{\bibinfo{journal}{Astropart.
  Phys.} \textbf{\bibinfo{volume}{108}}, \bibinfo{pages}{1}\bibinfo{year}{
  (\bibinfo{year}{2019})}}, ISSN \bibinfo{issn}{0927-6505}.

\bibitem[{\citenamefont{Aalseth et~al.}(2018)}]{aalsethDarkSide20k20Tonne2018}
\bibinfo{author}{\bibfnamefont{C.~E.} \bibnamefont{Aalseth}}
  \bibnamefont{et~al.} (\bibinfo{collaboration}{DarkSide Collaboration}),
  \href{https://doi.org/10.1140/epjp/i2018-11973-4}{\bibinfo{journal}{Eur.
  Phys. J. Plus} \textbf{\bibinfo{volume}{133}},
  \bibinfo{pages}{131}\bibinfo{year}{ (\bibinfo{year}{2018})}}, ISSN
  \bibinfo{issn}{2190-5444}.

\bibitem[{\citenamefont{Abi
  et~al.}(2018)}]{dunecollaborationDUNEFarDetector2018}
\bibinfo{author}{\bibfnamefont{B.}~\bibnamefont{Abi}}  \bibnamefont{et~al.}
  (\bibinfo{collaboration}{DUNE Collaboration}),
  \href{https://arxiv.org/abs/1807.10334}{\bibinfo{journal}{arXiv}:1807.10334\bibinfo{year}{
  (\bibinfo{year}{2018})}}.

\bibitem[{\citenamefont{Adhikari
  et~al.}(2023)}]{adhikariPrecisionMeasurementSpecific2023a}
\bibinfo{author}{\bibfnamefont{P.}~\bibnamefont{Adhikari}}
  \bibnamefont{et~al.} (\bibinfo{collaboration}{DEAP Collaboration}),
  \href{https://doi.org/10.1140/epjc/s10052-023-11678-6}{\bibinfo{journal}{Eur.
  Phys. J. C} \textbf{\bibinfo{volume}{83}},
  \bibinfo{pages}{642}\bibinfo{year}{ (\bibinfo{year}{2023})}}, ISSN
  \bibinfo{issn}{1434-6052}.

\bibitem[{\citenamefont{Ajaj
  et~al.}(2019{\natexlab{a}})}]{deapcollaborationElectromagneticBackgroundsPotassium422019}
\bibinfo{author}{\bibfnamefont{R.}~\bibnamefont{Ajaj}}  \bibnamefont{et~al.}
  (\bibinfo{collaboration}{DEAP Collaboration}),
  \href{https://doi.org/10.1103/PhysRevD.100.072009}{\bibinfo{journal}{Phys.
  Rev. D} \textbf{\bibinfo{volume}{100}},
  \bibinfo{pages}{072009}\bibinfo{year}{
  (\bibinfo{year}{2019}{\natexlab{a}})}}.

\bibitem[{\citenamefont{Adhikari
  et~al.}(2020{\natexlab{a}})}]{adhikariLiquidargonScintillationPulseshape2020}
\bibinfo{author}{\bibfnamefont{P.}~\bibnamefont{Adhikari}}
  \bibnamefont{et~al.} (\bibinfo{collaboration}{DEAP Collaboration}),
  \href{https://doi.org/10.1140/epjc/s10052-020-7789-x}{\bibinfo{journal}{Eur.
  Phys. J. C} \textbf{\bibinfo{volume}{80}},
  \bibinfo{pages}{303}\bibinfo{year}{ (\bibinfo{year}{2020}{\natexlab{a}})}},
  ISSN \bibinfo{issn}{1434-6052}.

\bibitem[{\citenamefont{Adhikari
  et~al.}(2021)}]{adhikariPulseshapeDiscriminationLowenergy2021}
\bibinfo{author}{\bibfnamefont{P.}~\bibnamefont{Adhikari}}
  \bibnamefont{et~al.} (\bibinfo{collaboration}{DEAP Collaboration}),
  \href{https://doi.org/10.1140/epjc/s10052-021-09514-w}{\bibinfo{journal}{Eur.
  Phys. J. C} \textbf{\bibinfo{volume}{81}},
  \bibinfo{pages}{823}\bibinfo{year}{ (\bibinfo{year}{2021})}}, ISSN
  \bibinfo{issn}{1434-6052}.

\bibitem[{\citenamefont{Ajaj
  et~al.}(2019{\natexlab{b}})}]{deapcollaborationSearchDarkMatter2019}
\bibinfo{author}{\bibfnamefont{R.}~\bibnamefont{Ajaj}}  \bibnamefont{et~al.}
  (\bibinfo{collaboration}{DEAP Collaboration}),
  \href{https://doi.org/10.1103/PhysRevD.100.022004}{\bibinfo{journal}{Phys.
  Rev. D} \textbf{\bibinfo{volume}{100}},
  \bibinfo{pages}{022004}\bibinfo{year}{
  (\bibinfo{year}{2019}{\natexlab{b}})}}.

\bibitem[{\citenamefont{Adhikari
  et~al.}(2020{\natexlab{b}})}]{deapcollaborationConstraintsDarkMatternucleon2020}
\bibinfo{author}{\bibfnamefont{P.}~\bibnamefont{Adhikari}}
  \bibnamefont{et~al.} (\bibinfo{collaboration}{DEAP Collaboration}),
  \href{https://doi.org/10.1103/PhysRevD.102.082001}{\bibinfo{journal}{Phys.
  Rev. D} \textbf{\bibinfo{volume}{102}},
  \bibinfo{pages}{082001}\bibinfo{year}{
  (\bibinfo{year}{2020}{\natexlab{b}})}}.

\bibitem[{\citenamefont{Adhikari
  et~al.}(2022)}]{deapcollaborationFirstDirectDetection2022}
\bibinfo{author}{\bibfnamefont{P.}~\bibnamefont{Adhikari}}
  \bibnamefont{et~al.} (\bibinfo{collaboration}{DEAP Collaboration}),
  \href{https://doi.org/10.1103/PhysRevLett.128.011801}{\bibinfo{journal}{Phys.
  Rev. Lett.} \textbf{\bibinfo{volume}{128}},
  \bibinfo{pages}{011801}\bibinfo{year}{ (\bibinfo{year}{2022})}}.

\bibitem[{\citenamefont{Heindl
  et~al.}(2010)}]{heindlScintillationLiquidArgon2010}
\bibinfo{author}{\bibfnamefont{T.}~\bibnamefont{Heindl}}  \bibnamefont{et~al.},
   \href{https://doi.org/10.1209/0295-5075/91/62002}{\bibinfo{journal}{Europhys.
  Lett.} \textbf{\bibinfo{volume}{91}}, \bibinfo{pages}{62002}\bibinfo{year}{
  (\bibinfo{year}{2010})}}.

\bibitem[{\citenamefont{Grace et~al.}(2015)\citenamefont{Grace, Butcher,
  Monroe, and Nikkel}}]{graceIndexRefractionRayleigh2015}
\bibinfo{author}{\bibfnamefont{E.}~\bibnamefont{Grace}},
  \bibinfo{author}{\bibfnamefont{A.}~\bibnamefont{Butcher}},
  \bibinfo{author}{\bibfnamefont{J.}~\bibnamefont{Monroe}},  \bibnamefont{and}
  \bibinfo{author}{\bibfnamefont{J.~A.} \bibnamefont{Nikkel}},
  \href{https://arxiv.org/abs/1502.04213}{\bibinfo{journal}{arXiv}:1502.04213\bibinfo{year}{
  (\bibinfo{year}{2015})}}.

\bibitem[{\citenamefont{Babicz
  et~al.}(2020)}]{babiczMeasurementGroupVelocity2020}
\bibinfo{author}{\bibfnamefont{M.}~\bibnamefont{Babicz}}  \bibnamefont{et~al.},
   \href{https://doi.org/10.1088/1748-0221/15/09/P09009}{\bibinfo{journal}{J.
  Instrum.} \textbf{\bibinfo{volume}{15}},
  \bibinfo{pages}{P09009}\bibinfo{year}{ (\bibinfo{year}{2020})}}.

\bibitem[{\citenamefont{Lane and
  Kuppermann}(2003)}]{laneArgonResonanceLine2003}
\bibinfo{author}{\bibfnamefont{A.~L.} \bibnamefont{Lane}} \bibnamefont{and}
  \bibinfo{author}{\bibfnamefont{A.}~\bibnamefont{Kuppermann}},
  \href{https://doi.org/10.1063/1.1683089}{\bibinfo{journal}{Rev. Sci.
  Instrum.} \textbf{\bibinfo{volume}{39}}, \bibinfo{pages}{126}\bibinfo{year}{
  (\bibinfo{year}{2003})}}.

\bibitem[{\citenamefont{Arai et~al.}(1978)\citenamefont{Arai, Oka, Kogoma, and
  Imamura}}]{araiInfraredAbsorptionsNeon1978}
\bibinfo{author}{\bibfnamefont{S.}~\bibnamefont{Arai}},
  \bibinfo{author}{\bibfnamefont{T.}~\bibnamefont{Oka}},
  \bibinfo{author}{\bibfnamefont{M.}~\bibnamefont{Kogoma}},  \bibnamefont{and}
  \bibinfo{author}{\bibfnamefont{M.}~\bibnamefont{Imamura}},
  \href{https://doi.org/10.1063/1.435565}{\bibinfo{journal}{J. Chem. Phys.}
  \textbf{\bibinfo{volume}{68}}, \bibinfo{pages}{4595}\bibinfo{year}{
  (\bibinfo{year}{1978})}}.

\bibitem[{\citenamefont{Landau et~al.}(1984)\citenamefont{Landau, Lifshittz,
  and Pitaevski}}]{landau1984electrodynamics}
\bibinfo{author}{\bibfnamefont{L.}~\bibnamefont{Landau}},
  \bibinfo{author}{\bibfnamefont{E.}~\bibnamefont{Lifshittz}},
  \bibnamefont{and}
  \bibinfo{author}{\bibfnamefont{L.}~\bibnamefont{Pitaevski}},
  \href{https://www.sciencedirect.com/book/9780080302751/electrodynamics-of-continuous-media}{\emph{\bibinfo{title}{Electrodynamics
  of Continuous Media, 2nd, rev. and enl.
  ed}}}\href{https://www.sciencedirect.com/book/9780080302751/electrodynamics-of-continuous-media}{,
  \bibinfo{publisher}{Institute of Physical Problems} (\bibinfo{address}{USSR
  Academy of Sciences} (\bibinfo{year}{1984})}, ISBN
  \bibinfo{isbn}{978-0-08-030275-1}.

\bibitem[{\citenamefont{Seidel et~al.}(2002)\citenamefont{Seidel, Lanou, and
  Yao}}]{seidelRayleighScatteringRaregas2002}
\bibinfo{author}{\bibfnamefont{G.~M.} \bibnamefont{Seidel}},
  \bibinfo{author}{\bibfnamefont{R.~E.} \bibnamefont{Lanou}},
  \bibnamefont{and} \bibinfo{author}{\bibfnamefont{W.}~\bibnamefont{Yao}},
  \href{https://doi.org/10.1016/S0168-9002(02)00890-2}{\bibinfo{journal}{Nucl.
  Instrum. Methods Phys. Res. A} \textbf{\bibinfo{volume}{489}},
  \bibinfo{pages}{189}\bibinfo{year}{ (\bibinfo{year}{2002})}}.

\bibitem[{\citenamefont{Lorentz}(1881)}]{lorentzUeberAnwendungSatzes1881}
\bibinfo{author}{\bibfnamefont{H.~A.} \bibnamefont{Lorentz}},
  \href{https://doi.org/10.1002/andp.18812480110}{\bibinfo{journal}{Annalen der
  Physik} \textbf{\bibinfo{volume}{248}}, \bibinfo{pages}{127}\bibinfo{year}{
  (\bibinfo{year}{1881})}}.

\bibitem[{\citenamefont{Clausius}(1879)}]{clausiusMechanischeBehandlungElectricitaet1879}
\bibinfo{author}{\bibfnamefont{R.}~\bibnamefont{Clausius}},
  \href{http://link.springer.com/10.1007/978-3-663-20232-5}{\emph{\bibinfo{title}{{Die
  Mechanische Behandlung der
  Electricit\"at}}}}\href{http://link.springer.com/10.1007/978-3-663-20232-5}{,
   \bibinfo{publisher}{{Vieweg+Teubner Verlag}} (\bibinfo{address}{{Wiesbaden}}
  (\bibinfo{year}{1879})}, ISBN \bibinfo{isbn}{978-3-663-19891-8
  978-3-663-20232-5}.

\bibitem[{\citenamefont{Rabinovich et~al.}(1987)\citenamefont{Rabinovich,
  Vasserman, Nedostup, and
  Veksler}}]{rabinovichThermophysicalPropertiesNeon1987}
\bibinfo{author}{\bibfnamefont{V.~A.} \bibnamefont{Rabinovich}},
  \bibinfo{author}{\bibfnamefont{A.~A.} \bibnamefont{Vasserman}},
  \bibinfo{author}{\bibfnamefont{V.~I.} \bibnamefont{Nedostup}},
  \bibnamefont{and} \bibinfo{author}{\bibfnamefont{L.~S.}
  \bibnamefont{Veksler}},
  \href{https://ui.adsabs.harvard.edu/abs/1988wdch...10.....R}{\emph{\bibinfo{title}{Thermophysical
  Properties of Neon, Argon, Krypton, and Xenon}}},
  vol.~\bibinfo{volume}{10}\href{https://ui.adsabs.harvard.edu/abs/1988wdch...10.....R}{,
  \bibinfo{publisher}{{Hemisphere Publishing Corporation, NY, NY}}
  (\bibinfo{year}{1987})}.

\bibitem[{\citenamefont{Lemmon}(2010)}]{ctx43101231760006531}
\bibinfo{author}{\bibfnamefont{E.~W.} \bibnamefont{Lemmon}},
  \href{https://webbook.nist.gov/chemistry/fluid/}{\bibinfo{journal}{NIST
  Chemistry WebBook}, \textbf{\bibinfo{volume}{1}},
  \bibinfo{pages}{1}\bibinfo{year}{ (\bibinfo{year}{2010})}}.

\bibitem[{\citenamefont{Sinnock and
  Smith}(1969)}]{sinnockRefractiveIndicesCondensed1969}
\bibinfo{author}{\bibfnamefont{A.~C.} \bibnamefont{Sinnock}} \bibnamefont{and}
  \bibinfo{author}{\bibfnamefont{B.~L.} \bibnamefont{Smith}},
  \href{https://doi.org/10.1103/PhysRev.181.1297}{\bibinfo{journal}{Phys. Rev.}
  \textbf{\bibinfo{volume}{181}}, \bibinfo{pages}{1297}\bibinfo{year}{
  (\bibinfo{year}{1969})}}.

\bibitem[{\citenamefont{Szydagis
  et~al.}(2021{\natexlab{a}})}]{szydagisReviewBasicEnergy2021}
\bibinfo{author}{\bibfnamefont{M.}~\bibnamefont{Szydagis}}
  \bibnamefont{et~al.} (\bibinfo{collaboration}{NEST Collaboration}),
  \href{https://doi.org/10.3390/instruments5010013}{\bibinfo{journal}{Instrum.}
  \textbf{\bibinfo{volume}{5}}, \bibinfo{pages}{13}\bibinfo{year}{
  (\bibinfo{year}{2021}{\natexlab{a}})}}.

\bibitem[{\citenamefont{{SCENE Collaboration}
  et~al.}(2013)}]{scenecollaborationObservationDependenceDrift2013}
\bibinfo{author}{\bibnamefont{{SCENE Collaboration}}}  \bibnamefont{et~al.}
  (\bibinfo{collaboration}{SCENE Collaboration}),
  \href{https://doi.org/10.1103/PhysRevD.88.092006}{\bibinfo{journal}{Phys.
  Rev. D} \textbf{\bibinfo{volume}{88}}, \bibinfo{pages}{092006}\bibinfo{year}{
  (\bibinfo{year}{2013})}}.

\bibitem[{\citenamefont{Cao
  et~al.}(2015)}]{scenecollaborationMeasurementScintillationIonization2015}
\bibinfo{author}{\bibfnamefont{H.}~\bibnamefont{Cao}}  \bibnamefont{et~al.}
  (\bibinfo{collaboration}{SCENE Collaboration}),
  \href{https://doi.org/10.1103/PhysRevD.91.092007}{\bibinfo{journal}{Phys.
  Rev. D} \textbf{\bibinfo{volume}{91}}, \bibinfo{pages}{092007}\bibinfo{year}{
  (\bibinfo{year}{2015})}}.

\bibitem[{\citenamefont{Agnes et~al.}(2018)}]{agnesMeasurementLiquidArgon2018}
\bibinfo{author}{\bibfnamefont{P.}~\bibnamefont{Agnes}}  \bibnamefont{et~al.}
  (\bibinfo{collaboration}{ARIS Collaboration}),
  \href{https://doi.org/10.1103/PhysRevD.97.112005}{\bibinfo{journal}{Phys.
  Rev. D} \textbf{\bibinfo{volume}{97}}, \bibinfo{pages}{112005}\bibinfo{year}{
  (\bibinfo{year}{2018})}}.

\bibitem[{\citenamefont{Bolton et~al.}(2018)}]{rat}
\bibinfo{author}{\bibfnamefont{T.}~\bibnamefont{Bolton}}  \bibnamefont{et~al.},
  \emph{\bibinfo{title}{{RAT (is an Analysis Tool) User's Guide}}}
  (\bibinfo{year}{2018}), \urlprefix\url{https://rat.readthedocs.io}.

\bibitem[{\citenamefont{Agostinelli et~al.}(2003)}]{GEANT4}
\bibinfo{author}{\bibfnamefont{S.}~\bibnamefont{Agostinelli}}
  \bibnamefont{et~al.},
  \href{https://doi.org/10.1016/S0168-9002(03)01368-8}{\bibinfo{journal}{Nucl.
  Instrum. Methods Phys. Res. A} \textbf{\bibinfo{volume}{506}},
  \bibinfo{pages}{250}\bibinfo{year}{ (\bibinfo{year}{2003})}}.

\bibitem[{\citenamefont{Safarabadi}(2023)}]{sinaThesis}
\bibinfo{author}{\bibfnamefont{S.}~\bibnamefont{Safarabadi}}, Ph.D. thesis,
  \bibinfo{school}{University of Alberta} (\bibinfo{year}{2023}).

\bibitem[{\citenamefont{Szydagis
  et~al.}(2021{\natexlab{b}})}]{szydagisInvestigatingXENON1TLowEnergy2021}
\bibinfo{author}{\bibfnamefont{M.}~\bibnamefont{Szydagis}}
  \bibnamefont{et~al.} (\bibinfo{collaboration}{NEST Collaboration}),
  \href{https://doi.org/10.1103/PhysRevD.103.012002}{\bibinfo{journal}{Phys.
  Rev. D} \textbf{\bibinfo{volume}{103}},
  \bibinfo{pages}{012002}\bibinfo{year}{
  (\bibinfo{year}{2021}{\natexlab{b}})}}.

\bibitem[{\citenamefont{Joshi
  et~al.}(2014)}]{joshiFirstMeasurementIonization2014}
\bibinfo{author}{\bibfnamefont{T.~H.} \bibnamefont{Joshi}}
  \bibnamefont{et~al.},
  \href{https://doi.org/10.1103/PhysRevLett.112.171303}{\bibinfo{journal}{Physical
  Review Letters} \textbf{\bibinfo{volume}{112}},
  \bibinfo{pages}{171303}\bibinfo{year}{ (\bibinfo{year}{2014})}}.

\end{thebibliography}
\end{document}